\documentclass[usenatbib]{mnras}
\usepackage{amsmath}
\usepackage{xcolor}
\usepackage{graphicx}
\usepackage{amssymb}

\title[\textit{periapsis}: An Orbital Fitting Package]{\textit{periapsis}: A Comprehensive Orbital Fitting Package}

\author[E. Kerr et al.]{
Evan Kerr,$^{1}$\thanks{E-mail: evan.kerr@utah.edu}
Matthew Whitaker,$^{1}$
Anil Seth,$^{1}$
Jay Strader,$^{2}$
Maximilian H\"{a}berle,$^{3}$
\newauthor
and Sebastian Kamann$^{4}$
\\
$^{1}$Department of Physics and Astronomy, University of Utah, Salt Lake City, UT 84112, USA\\
$^{2}$Michigan State University, USA\\
$^{3}$European Southern Observatory, Germany\\
$^{4}$Liverpool John Moores University, UK
}

\date{Accepted XXX. Received YYY; in original form ZZZ}

\pubyear{2026}

\begin{document}
\label{firstpage}
\pagerange{\pageref{firstpage}--\pageref{lastpage}}
\maketitle

\begin{abstract}
We introduce a new Python package \texttt{periapsis}, which computes  posterior distributions for a variety of orbital parameters given astrometric and/or radial velocity data. \texttt{periapsis} not only returns posterior distributions, but also a variety of statistics assessing convergence and fit metrics as well as plotting capabilities for visualizing the orbital systems. In this paper, we describe our various methods for fitting Keplerian orbits to both astrometric and radial velocity data. We then show the breadth of binary systems that \texttt{periapsis} can characterize by comparing our computed parameters with known orbital systems. Finally, we discuss how \texttt{periapsis} will be a helpful tool for future data releases like \textit{Gaia} DR4 and \textit{Roman}.   
\end{abstract}


\section{Introduction}
\label{intro}

A large fraction of stars along the main sequence are believed to form and reside in binary or higher-order systems \citep{Offner2023,Moe2017}. These systems are rich with interesting astrophysical phenomena enabling us to learn more about stellar evolution \citep{Moe2017}, compact object formation and populations \citep{andrews_weighing_2019,Muller-Horn2026}, planetary systems \citep{Butler2006}, progenitors to gravitational wave sources \citep{Abbott2016}, and much more. The last few decades of astronomy have seen a dramatic increase in the number of spectroscopic surveys, providing a large number of radial velocity measurements on stars across the Milky Way \citep{GALAH2015,APOGEE2017}, and with the advent of \textit{Gaia}, astrometry for over a billion sources has been recorded \citep{GaiaDR3_2023,GaiaDR3Halbwachs}. These datasets have led to the discovery of many interesting stellar, planetary, and compact object companions, greatly increasing our knowledge on their populations \citep{Muller-Horn-WR-2026,Butler2006,ElBadry2023_BH1,ElBadry2023_BH2,GaiaBH3_2024,Whitaker2026}. Being able to accurately characterize these systems and their uncertainties is important to further our understanding of these populations and phenomena. 

Many different approaches have been taken to characterize binary systems, with Bayesian inference being highly favored to obtain posterior distributions of orbital parameters that encapsulate their uncertainties \citep{Lucy2018}. Different ways of doing this have been presented in a variety of orbital fitting packages and custom made codes, made to fit astrometry, radial velocity, or joint measurements \citep{TheJoker2017,orbitize2020,El-Badry2024}. However, it can be difficult to find a package that works for a variety of observational scenarios and that is easy and intuitive to use. 

We here present our comprehensive orbital fitting package \texttt{periapsis}\footnote{\url{https://github.com/Evank2004/periapsis}}, which is capable of fitting orbits to a wide variety of observational data, whether that be 2D astrometry, scanning-plane astrometry (like \textit{Gaia}), radial velocities, or any combination of the three. The package can also accommodate the case where both components of the binary system have been observed, performing a joint fit on both. The package has been designed to fully characterize a binary system with a minimal amount of manual tuning done by the user. This ``plug and play" design allows for easy analyses of interesting systems, where users can obtain detailed orbital constraints on their system with a minimal amount of code. The package can return a large variety of statistics on the behavior of the sampler and the parameter values and uncertainties. To visualize the orbital fit and the sampler behavior, multiple plotting options are included.   

The rest of the paper is presented as follows. Section \ref{sec:Keplerian Models} presents the orbital models used to calculate model positions and velocities, as well as the formulation for other parameters and quantities that are able to be fit for and returned. Section \ref{sec:fitting algorithm} then describes the full fitting algorithm, going over the full pipeline from data input to results. We then show the effectiveness of \texttt{periapsis} by fitting known orbital  systems in Section \ref{sec:experiments}. We conclude with a discussion in Section \ref{sec:discussion} featuring tips on how to obtain a quick orbital solution and outlook on future uses of the package.

\section{Keplerian Models}
\label{sec:Keplerian Models}
Here we describe how we fit a Keplerian orbital model to multi-epoch astrometry data, radial velocity measurements, and joint datasets combining the two. While our model uses a primary set of fitting parameters, $\texttt{periapsis}$ supports sampling for many different orbital parameterizations, which can then be transformed into the inputs required by our model. A full list of supported parameters can be found in Table  \ref{tab:periapsis_params}, while any relationship between parameters not presented in this section can be found in Appendix \ref{app:sec1}. We also detail the linearization of the Keplerian model, separating the non-linear parameters (period $P$, eccentricity $e$, and time of periastron passage $T_p$) from the remaining linear ones (Section~2.3). The remaining parameters can then be solved analytically with our dynamic, observation-defined matrix. Then, Section \ref{subsec:additional_params} describes all the additional parameters that can be incorporated to our models for sampling. 

Unless specified otherwise, parameters presented with a subscript $1$ correspond to the more luminous companion, while a subscript of $2$ corresponds to the less luminous or unseen companion. In practice, the assignments of system $1$ and $2$ are arbitrary, and the user must be sure to consistently assign them in the data class (Section \ref{data and units}). 

\subsection{Astrometry}
\label{astrometry_model}

Astrometric observations are assumed to be defined relative to a reference point ($\alpha_0,\delta_0$) with the Right Ascension offset $\Delta\alpha^* = \Delta\alpha\cos{\delta_0}$ (for small offsets) and Declination offset $\Delta\delta$ of the source. The path of the source will show periodic astrometric variations when undergoing Keplerian orbital motion if there is more than one body in the system. The motion of the observer induces additional changes. \texttt{periapsis} is capable of modeling this motion for resolved and unresolved binary systems. The displacement of a source due to parallax $\varpi$, proper motion $\mu$, and orbital motion is given, for each observational epoch $t_k$, as

\begin{equation}
\label{eq:alpha}
    \Delta\alpha^*= B X_k + GY_k + \Delta\alpha^*_0 + \mu_{\alpha^*} \Delta t + \Pi_{\alpha^*}\varpi
\end{equation}
\begin{equation}
\label{eq:delta}
    \Delta\delta = A X_k + FY_k + \Delta\delta_0 + \mu_\delta \Delta t + \Pi_\delta \varpi 
\end{equation}

where $\mu_{\alpha^*}$ and $\mu_\delta$ give the proper motions, and $\Delta t = (t_k-T_{\text{epoch}})$, where $T_{\text{epoch}}$ is a conveniently defined reference epoch. The terms $\Pi_{\alpha^*}$ and $\Pi_{\delta}$ are time dependent parallax factors, which project the Earth's orbit about the barycentre of the Solar System onto the sky plane, normalized to unity (\cite{McArthur2011}, see equations $53$ \& $54$ in \citet{Wright2009}). $A,B,F,G$ are the Thiele-Innes elements, transformations of four of the Campbell elements: semi-major axis $a$, inclination $i$, argument of periastron $\omega$ and longitude of the nodes $\Omega$ (see Appendix \ref{app:Thiele-Innes}).

We define the argument of periastron  $\omega$ in the standard astrometric convention \citep{Lucy2014}, where it is measured from the ascending node in the direction of orbital motion. We define system-specific components $\omega_1$ and $\omega_2$ whose values correspond to the respective component's orbit with respect to the barycentre. This ensures our definition is fully consistent between astrometry and radial velocity models. When fitting pure astrometry data, the line of sight orientation determined by the sign of the inclination $i$ is unconstrained. This causes the astrometric fit to be invariant under $+\pi$ phase shifts of $\omega$ and $\Omega$, which can only be fully constrained with additional radial velocity data. The longitude of the ascending node $\Omega_{\text{asc}}$ can also be fully constrained from joint fits. 

The terms $X_k$ and $Y_k$ represent the elliptical coordinates in the orbital plane per epoch, which depend on the non-linear parameters $P$, $e$, $T_p$:
\begin{align}
    X_k = \cos{E} -e\\
    Y_k = \sqrt{1-e^2}\sin{E}
\end{align}
where $E$ is the eccentric anomaly obtained by solving Kepler's equation,
\begin{equation}
    \frac{2\pi}{P}(t_k-T_p) = E -e\sin{E}.
\end{equation}

While relative sky-plane astrometry directly measures an instantaneous 2D positional offset, absolute astrometry scanning space missions like \textit{Gaia} precisely measure the abscissa of the star along a scan direction while the orthogonal axis is determined with much lower precision. For each observation epoch $t_k$, \textit{Gaia} records an along scan measurement $w_k$ projected along the instrument scan angle $\psi_k$. To evaluate the likelihood for this model, we project $\Delta\alpha^*$,$\Delta\delta$ onto the scan angle: 
\begin{equation}
\begin{aligned}
    w_{\text{model}}(t_k) &= [B X_k + G Y_k + \Delta\alpha^*_0 + \mu_{\alpha^*} \Delta t]\sin{\psi_k} \\
    &\quad + [A X_k + F Y_k + \Delta\delta_0 + \mu_\delta \Delta t]\cos{\psi_k}\\
    &+ \Pi_k \varpi.
\end{aligned}
\end{equation}
Where $\Pi_k$ is the \textit{Gaia} reported along-scan parallax factor per epoch, which is related to the parallax factors in Equations \ref{eq:alpha} and \ref{eq:delta} by $\Pi_k = \Pi_{\alpha^*} \sin{\psi_k} + \Pi_\delta \cos{\psi_k}$. 

\subsection{Radial Velocities}
\label{rv_model}
RV measurements are expected as a time series of the barycentric radial velocity $v$ at times $t_k$. For a system undergoing Keplerian motion, we model the line of sight (radial) velocity as
\begin{equation}
\label{eq:RV full model}
\begin{aligned}
    v_{\text{model}}(t_k)= K (\cos{(\nu+\omega)} + e\cos{\omega}) + \gamma.
\end{aligned}   
\end{equation}
Here, $K$ is the radial velocity semi-amplitude, $\gamma$ is the barycentric systemic velocity and $\nu$ is the true anomaly given by
\begin{equation}
\begin{aligned}
\nu = 2 \arctan{\sqrt{\frac{1+e}{1-e}}\tan{\frac{E}{2}}}.
\end{aligned}
\end{equation}

To separate the linear parameters, we use a slightly different model where
\begin{equation}
\label{eq:linear_rv}
\begin{aligned}
v_\text{model}(t_k) = h(\cos{\nu}+e) + c\sin{\nu} + \gamma.
\end{aligned}
\end{equation}
Here $c$ and $h$ are constants closely related to the Thiele-Innes elements $C$ and $H$ \citep{Wright2009}, and are given by 
\begin{equation}
    h = K\cos{\omega},
\end{equation}
\begin{equation}
\label{eq:c}
    c = - K \sin{\omega}.
\end{equation}

Throughout this paper, $c$ always refers to the definition given in Equation \ref{eq:c}, not the speed of light. In the case that both systems are observed or enough information is obtained to define both, Equation \ref{eq:RV full model} or \ref{eq:linear_rv} is solved for both systems $1$ and $2$ with their unique components ($K_1,K_2,\omega_1,\omega_2$).

\subsection{Linearization}
\label{linearization}
The parameter vectors corresponding to the models presented in Sections \ref{astrometry_model} and \ref{rv_model} can be sampled in full by \texttt{periapsis}. However, the linear parameters can also be separated from the non-linear ones, allowing us to sample just $P$, $e$, and $T_p$. This method is widely used, as it reduces the dimensionality of the sampling space, leading to faster convergence and improved computational efficiency \citep{Wright2009,Lucy2014,El-Badry2024}. 

 We follow the linearization procedure from \citet{El-Badry2024}, but generalize it to the non-Gaia case, where astrometry is defined as $\Delta \alpha^*$ and $\Delta \delta$ as above, and incorporating RVs.  Given a proposed $P$, $e$ and $T_p$, the linear parameters can be solved analytically via linear least squares.  We start by writing:
\begin{equation}
    \vec{x} = \mathbf{M}\vec{\mu},
\end{equation}
where $\vec{x}$ is the vector of observational data, $\mathbf{M}$ is the design matrix, and $\vec{\mu}$ is the linear parameter vector. Consider a case where both relative astrometry and RVs are available. The data vector $\vec{x}$ and linear parameter vector $\vec{\mu}$ take the form:
\begin{equation}
\begin{aligned}
\vec{x} &= (\Delta\alpha^*, \Delta\delta, v_{\text{RV}})^T, \\
\vec{\mu} &= (\Delta\alpha_0, \mu_{\alpha^*}, \varpi, \Delta\delta_0, \mu_\delta,B, G, A, F,  \gamma, h, c)^T,
\end{aligned}
\end{equation}
and the dynamic design matrix $\mathbf{M}$ is constructed as a block matrix:
\begin{equation}
\label{eq:matrix_M}
\mathbf{M} = 
\begin{pmatrix} 
\mathbf{M}_{\text{astro}} & {0} \\ 
{0} & \mathbf{M}_{\text{RV}} 
\end{pmatrix},
\end{equation}
where the $\mathbf{M}_{\text{astro}}$ astrometric sub-matrix operating on $(\Delta\alpha_0, \mu_{\alpha^*}, \varpi, \Delta\delta_0, \mu_\delta, B, G, A, F)^T$ is given by:
\begin{equation}
\mathbf{M}_{\text{astro}} = 
\begin{pmatrix}
{1} & \Delta {t} & \Pi_{\alpha^*} & {0} & {0} & {X_k} & {Y_k} & {0} & {0}
\\ {0} & {0} & \Pi_\delta & {1} & \Delta {t} & {0} & {0} & {X_k} & {Y_k}
\end{pmatrix},
\end{equation}
and the radial velocity sub-matrix operating on $(\gamma, h, c)^T$ is given by:
\begin{equation}
\mathbf{M}_{\text{RV}} = 
\begin{pmatrix}
{1} & \cos{\nu} + e & \sin{\nu}
\end{pmatrix}.
\end{equation}

In the case of 1D \textit{Gaia} astrometry, the astrometric rows are projected along the instrument scan angle, resulting in matrix $\mathbf{A}$ given in Appendix A of \citet{El-Badry2024}. 

The maximum likelihood solution for the linear parameter vector $\vec{\mu}$ is then evaluated at each sampling step via weighted linear least squares:
\begin{equation}
    {\vec{\mu}} = (\mathbf{M}^T \mathbf{W} \mathbf{M})^{-1} \mathbf{M}^T \mathbf{W} \vec{x},
\end{equation}
where the observational uncertainties $\sigma_k$ are taken into account with the inverse covariance matrix $\mathbf{W} = \text{diag}(1/\sigma_k^2)$.

\subsection{Systemic Trends, Mass Functions, and Null Hypothesis Models}
\label{subsec:additional_params}
In addition to pure Keplerian orbital motion, \texttt{periapsis} can account for instrumental offsets, long-term RV trends, flux ratios, jitter, physical mass constraints, and null hypothesis fitting. Here, we present how each of these are incorporated into our framework. 

\subsubsection{Instrumental Offsets and Acceleration Trends}
\label{subsub:instr_offsets_trends}

When combining observations from multiple instruments, potential offsets between the instrumental reference frames can be accounted for. For each instrument provided, an instrumental offset term can be added linearly to the corresponding model. For example, if multiple instruments are used for radial velocity measurements, then an offset term for each telescope will be linearly added to Equation \ref{eq:RV full model} relative to a chosen reference frame.

Additionally, to account for acceleration caused by long-period companions, one can fit for a linear radial velocity trend $\dot{\gamma}$ \citep{Wright2009} by adding $\dot\gamma \Delta t$ to Equation \ref{eq:RV full model}. Astrometric acceleration can be accounted for through the parameters $\dot{\mu_{\alpha^*}}$, $\dot{\mu_\delta}$ by adding $\frac{1}{2} \dot{\mu_{\alpha^*}} \Delta t^2$, $\frac{1}{2} \dot{\mu_\delta} \Delta t^2$ to Equations \ref{eq:alpha} and \ref{eq:delta}, respectively. 

Both the instrumental offsets and the acceleration trend parameters can be incorporated with our linear model (Section \ref{linearization}) by simply including these terms in the parameter vector $\vec{\mu}$. Their corresponding elements in the sub-matrices ($\mathbf{M}_{\text{astro}}$ and $\mathbf{M}_{\text{RV}})$ are dynamically populated with $1$ for specified instrumental offsets, $\mathbf{M_{\text{RV}}}$ is populated with $\Delta t$ for $\dot{\gamma}$, and $\mathbf{M_{\text{astro}}}$ is populated with $\Delta t^2$ for the astrometric acceleration parameters.

\subsubsection{Flux Ratios}
\label{subsub:flux_ratios}

In the case that both bodies in an unresolved astrometric binary are luminous, the photocentre will no longer trace a single star, but will fall on a line between the two bodies. The photocentre semimajor axis is given as
\begin{equation}
\begin{aligned}
    \tilde{a} = a_1[1-\frac{f(1+q)}{q(1+f)}],
\end{aligned}
\end{equation}

where the mass ratio $q = M_2/M_1$, $a_1$ is the primary's semi-major axis, and $f=F_2/F_1$ is the flux ratio. $F_1$ and $F_2$ correspond to flux from the primary and secondary companions, respectively. The filter for the flux is provided by the user in the data class (Section \ref{data and units}), and a flux ratio can be fit for each filter in the data set. This is applied to our model by multiplying the photocentre factor $f_{\text{photo}}=[1-\frac{f(1+q)}{q(1+f)}]$ to the Thiele-Innes elements $A_1,B_1,F_1,G_1$. $f$ will be consistent with zero when the binary system contains a dark companion and systematic-free astrometry \citep{El-Badry2026}.

\subsubsection{Jitter}
\label{subsub:jitter}

Astrometry and radial velocity measurements may show higher variance than expected from instrumental uncertainties $\sigma_k$ alone due to unmodeled stellar activity and additional instrument systematics \citep{Baluev2009,Taaki2026}. To account for this, \texttt{periapsis} can fit for a jitter term $\sigma_{\text{jit}}$ by defining the total variance as
\begin{equation}
\begin{aligned}
    \sigma_{\text{tot}}^2 = \sigma_k^2 + \sigma_{\text{jit}}^2.
\end{aligned}
\end{equation}

The log-likelihood for $N$ observations is then modified to incorporate $\sigma_{\text{tot}}$ in place of $\sigma_k$:
\begin{equation}
\label{eq:lnlike}
\begin{aligned}
    \ln \mathcal{L} = -\frac{1}{2} \sum_{k=1}^{N}[(\frac{x_{\text{obs},k} - x_{\text{model},k}}{\sigma_{\text{tot},k}})^2 + \ln(2 \pi \sigma_{\text{tot},k}^2)],
\end{aligned}
\end{equation}
where the errors are assumed to be Gaussian and uncorrelated.

\subsubsection{Mass Constraints} 
Different constraints on the masses of the binary components can be made depending on the type of observation made. RV observations are inherently limited to measuring the minimum masses due to the unknown inclination angle $i$, while astrometry can provide constraints on the inclination directly.  

We infer dynamical mass estimates through Kepler's third law, combining the orbital period $P$ with the physical semi-major axis. This is done in the canonical unit system (Section \ref{data and units}) such that the gravitational constant $G=4\pi^2$. Angular astrometric semi-major axes $a_{\rm ang}$ are converted to physical units using the parallax, $a_{\rm phys}[\mathrm{AU}]=a_{\rm ang}[\mathrm{mas}]/\varpi[\mathrm{mas}]$.

For cases where both components of the system have been observed, the mass ratio $q$ can be determined with
\begin{equation}
\label{eq:mass_ratio}
 q = \frac{M_2}{M_1} = \frac{a_1}{a_2} = \frac{K_1}{K_2} 
\end{equation}.

Radial velocity data specifies the RV primary mass function 
\begin{equation}
\label{eq:rv_mass_func}
\begin{aligned}
    f_{\rm RV,1}
    (M)=\frac{P K_1^3}{2\pi G}(1-e^2)^{3/2} =\frac{M_2^3\sin^3 i}{(M_1+M_2)^2},
\end{aligned}
\end{equation}
which constrains the minimum mass of the secondary (for the case where $M_1 = 0$ and $i=\pi/2$, giving sin $i=1$). The analogous $f_{\rm RV,2}(M)$ is defined by $K_2$ and constrains the minimum mass of the primary. 

Assuming a visible primary and an invisible secondary, the astrometric mass function for the unseen companion is given as
\begin{equation}
\label{eq:astro_M2}
\begin{aligned}
    f_{\rm astro,1}(M)
    =
    \frac{a_1^3}{P^2} = {\frac{M_2^3}{(M_1+M_2)^2}}.
\end{aligned}
\end{equation}

If the primary mass $M_1$ is known \citep{andrews_weighing_2019}, the companion mass $M_2$ or its minimum $M_{2 ,\rm min }$ can be uniquely determined, for the astrometric and RV cases, respectively. Both Equations \ref{eq:astro_M2} and \ref{eq:rv_mass_func} are then solved numerically for each posterior sample to obtain $M_2$ and $M_{2,\rm min}$.


\subsubsection{Null Hypothesis Models}
\label{subsub:null_models}

To check whether a Keplerian orbit is truly the best descriptor of the observed data, one can compare the $\chi^2$ of the orbital fit with a single star null hypothesis fit. 

The single star model for astrometry is modeled as 
\begin{equation}
\label{eq:null_astro}
\begin{aligned}
    \Delta\alpha_{ss}^{*} &= \Delta\alpha_{0}^{*} + \mu_{\alpha^{*}}\Delta t + \Pi_{\alpha^{*}}\varpi \\
    \Delta\delta_{ss} &= \Delta\delta_{0} + \mu_{\delta}\Delta t + \Pi_{\delta}\varpi.
\end{aligned}
\end{equation}

While the single star model for \textit{Gaia} astrometry is
\begin{equation}
\label{eq:null_astro_gaia}
\begin{aligned}
    w_{ss}(t_k) ={}& \left[ \Delta\alpha^*_0 + \mu_{\alpha^*} \Delta t \right] \sin \psi_k \\
                  & + \left[ \Delta\delta_0 + \mu_\delta \Delta t \right] \cos \psi_k 
                  + \Pi_k \varpi.
\end{aligned}
\end{equation}

For radial velocities, the null model simply fits a constant systemic velocity $\gamma$. 

These models are all fit via linear least squares to obtain a quick estimate on the $\chi^2$ of the single star model. The difference $\Delta\chi^2$ between the single star fit and the full Keplerian fit is used to create $\chi^2$ parameter grids to obtain initial estimates for seeding the sampler (Section \ref{initial guess}) and reported at the end of fitting (Section \ref{Output}). 




\section{\texttt{periapsis} Fitting Algorithm}
\label{sec:fitting algorithm}

\begin{figure*}
    \centering
    \includegraphics[width=\textwidth]{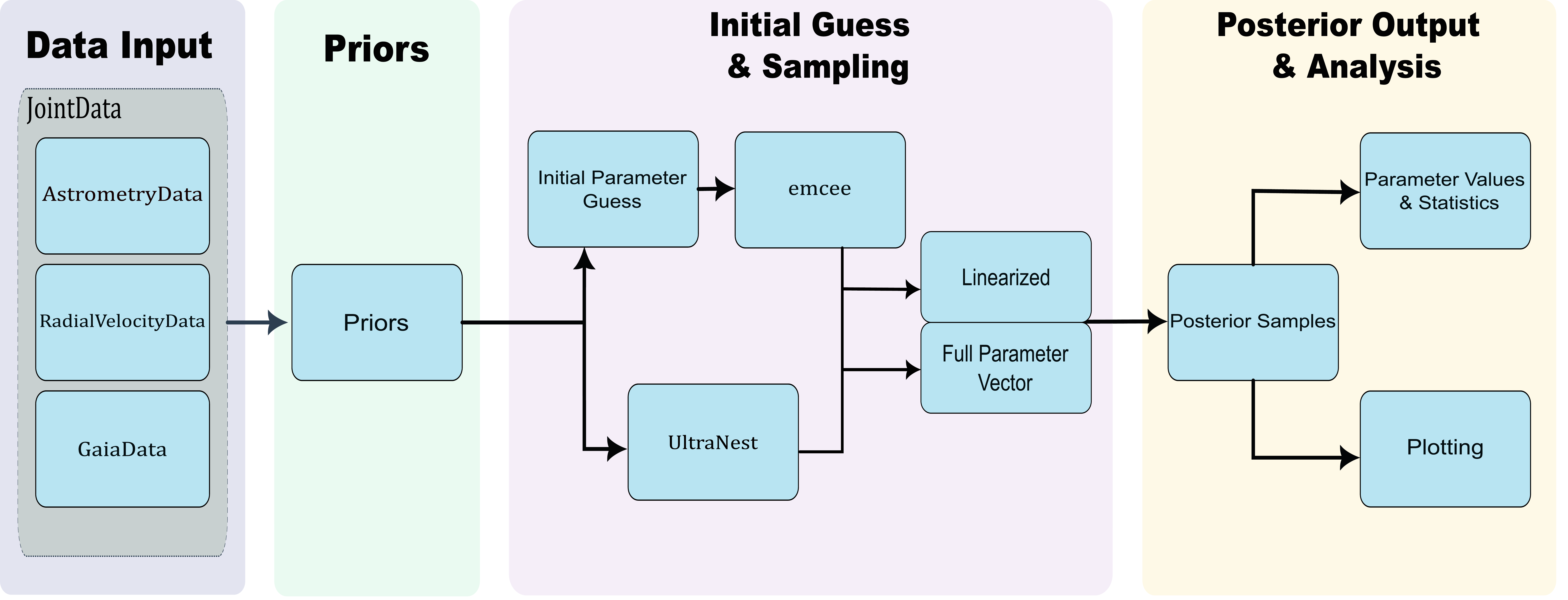}
    \caption{Flowchart of the \texttt{periapsis} orbital parameter estimation framework.}
    \label{fig:flowchart}
\end{figure*}

\texttt{periapsis} uses Bayesian inference to infer a posterior probability distribution for the orbital parameters and companion properties from multiple types of observations, such as 2D relative astrometry, \textit{Gaia} 1D scan-plane astrometry, radial velocities. It has been made to require minimal manual configuration by the user while providing a streamlined pipeline that handles unit normalization, parameter initialization, posterior sampling, statistical analysis, and scientific visualization.

As illustrated in Figure  \ref{fig:flowchart}, the package's workflow proceeds in the following steps: 
\begin{enumerate}
    \item \textbf{Data Input:} Observational measurements, errors and their corresponding units are ingested and converted into package's canonical unit system (Section \ref{data and units}).
    \item \textbf{Priors:} Users can define prior probability distributions over both sampled and derived physical parameters. Uniform, normal, beta, log-uniform, and log-normal distributions are directly supported (Section \ref{priors}).
    \item \textbf{Initial Guesses:} Sampling with \texttt{emcee} \citep{emcee} requires an initial seed to start exploration of the parameter space. Multiple initial fitting schemes are offered depending on the observational data supplied to efficiently locate a plausible global minima within the parameter space to seed the sampler (Section \ref{initial guess}).
    \item \textbf{Sampling:} To fit a Keplerian orbit to the input data, \texttt{periapsis} utilizes Bayesian posterior sampling to determine the orbital parameters. Two sampling methods are offered for Bayesian inference with either the Markov Chain Monte Carlo (MCMC) technique via \texttt{emcee} \citep{emcee}, or Nested Sampling via \texttt{UltraNest} \citep{Ultranest} (Section \ref{sampling}).
    \item \textbf{Output:} Once sampling is completed, \texttt{periapsis} returns the posterior distribution for sampled and derived physical parameters. It can then provide a variety of statistics discussing the fit properties and sampler convergence, as well as visualizations for the sampler characteristics and orbital fit (Section \ref{Output}).
\end{enumerate}

\subsection{Data and Unit Initialization}
\label{data and units}

Provided data and its corresponding units are input and ingested through the package's data classes \texttt{AstrometryData}, \texttt{RadialVelocityData} and \texttt{GaiaData}. \texttt{periapsis} uses \texttt{astropy.units} \citep{astropy:2013,astropy:2018,astropy:2022} to parse the supplied units and ensure that quantities within each dataset are dimensionally compatible. This allows users to provide data in any standard astronomical unit without manual conversion before fitting.  

The user must provide units for their data. All quantities are then converted into a canonical unit system upon ingestion. The assumed canonical units are years for time, astronomical units for linear distances, radians for angular quantities, $\mathrm{AU/yr}$ for linear velocities, and solar masses for mass. In the canonical unit system, the gravitational constant $G=4\pi^2$.

The dimensionality and original unit for each quantity is saved throughout the analysis, where the modeled and derived parameters are calculated in the canonical unit system, and output in the user specified units. A parallax or distance is needed to convert from angular astrometric quantities to physical linear quantities.

Users must specify which system component their data corresponds to in the data class by setting the system to either \textbf{`relative'}, \textbf{`1'} or \textbf{`2'}. This allows \texttt{periapsis} to correctly define parameters and model the motion for each individual component. The filter and instrument corresponding to each data can also be supplied in the data classes, which are then taken to make a variable for an instrumental offset parameter (Section \ref{subsub:instr_offsets_trends}) and flux ratio (Section \ref{subsub:flux_ratios}) if those parameters are desired to be fit. Any number of data classes can be constructed and put into the \texttt{JointData} class, which will then fit all supplied datasets with a joint likelihood. This allows for joint fits between radial velocity and astrometry measurements, multiple instruments, and enables a way to fit for both components of the binary system if both have been observed.

\subsection{Priors}
\label{priors}

Inputted priors are what define the parameters desired for sampling. The amount of parameters supplied must define enough parameters to model the provided data (Section \ref{astrometry_model}, \ref{rv_model}), or in the case of linearization (Section \ref{linearization}), only the non-linear parameters $P,e,T_p$ must have specified priors (or parameters that can transform to them). 

If no units are provided for prior values, they are expected to be input with the units used in the data class, meaning that if a user supplied their observation times in years, then a prior on period $P$ would be in units of years. The priors are then converted into the canonical unit system for internal calculations. Unit conversion is a linear transformation, which has no impact on sampling or convergence. This is because \texttt{emcee} is affine invariant \citep{emcee} and \texttt{UltraNest} maps the parameter space to a unit hypercube \citep{Ultranest}. 

\texttt{periapsis} offers supported prior probability distributions for both sampled and derived physical parameters. Standard uniform (\texttt{UniformPrior}), normal (\texttt{NormalPrior}) and beta (\texttt{BetaPrior}) distributions are offered, along with log counterparts (\texttt{LogUniformPrior} and \texttt{LogNormalPrior}). 

For derived parameters, we offer a \texttt{Bounds} prior class, which simply sets the bounds on a derived parameter. For example, if you know your data set does not contain any binary with a companion mass $> 50 M_\odot$, a bound is set on the derived companion mass $M_2$, meaning any sample that has an $M_2$ outside the bounds is rejected. The \texttt{FixedPrior} class is offered to allow users to fix parameters, like setting $e=0$ for known circular orbits. This is also the standard way to set a known primary mass $M_1$ and a  reference epoch $T_{\text{epoch}}$. 

\texttt{UltraNest} requires priors to be transformed to a unit hypercube. \texttt{periapsis} handles this by automatically transforming the provided priors, allowing users to seamlessly switch between \texttt{emcee} and \texttt{UltraNest} with a single set of priors. 

\subsection{Determining Initial Guesses}
\label{initial guess}
To implement \texttt{emcee}, one must first provide an initial guess to the sampler. The initial estimated parameters can significantly impact the performance and accuracy of MCMC.  Initial guesses far from the correct solution can result in posteriors getting stuck in local minima. We find that if the sampler has a decent estimate on the period of the orbital system, the time to convergence is significantly decreased and getting trapped in local minima is less likely. We introduce multiple initial fitting schemes that are unique to the types of data that one is performing an orbital fit on. 

In this section, we first detail the specific periodogram framework used to find an initial period estimate on different data types. We then describe how we obtain an initial estimate for the remaining parameters to provide a robust walker seed for \texttt{emcee}.

\subsubsection{2D Astrometry}
\label{subsub:initial_2d_astro}
To determine an initial period estimate $P_{\text{init}}$ from irregularly sampled time series, the package utilizes the Lomb-Scargle (LS) \citep{Lomb,Scargle1982,VanderPlas} implementation in \texttt{astropy} to evaluate power spectra along both orthogonal axes. The spectra are then added into a combined spectrum, where a specified number of peaks from the combined power yields the primary candidate frequencies, whose inverse are the candidate $P_{\text{init}}$ estimates. The amount of peaks selected for candidate frequencies by default is one, but the user is free to specify any number of peaks in the \texttt{emcee} fitter classes with the \texttt{npeaks} argument (Section \ref{subsubsec:emcee}).   

The sensitivity of the LS periodogram can degrade in high eccentricity regimes due to transfer of spectral power from the fundamental frequency into higher order harmonics. However, for astrometry the power fraction in the fundamental frequency remains above $80\%$ \citep{Delisle2022}. While other methods have been developed to model the periodicity in high eccentricity regimes \citep{zucker_detection_2019,Delisle2022}, the high computational efficiency of the LS algorithm and the minor fractional power loss makes the LS periodogram well suited for seeding the initialization pipeline.

The LS fit also provides a heuristic estimate on the semi-major axis $a_{init}$. By taking the best-fitting trigonometric amplitudes associated with the primary candidate frequency along each axis ($A_{\Delta\alpha^*}, A_{\Delta\delta}$), the spatial extent of the orbit can be estimated by combining the amplitudes in quadrature, $a_{init}$ = $\sqrt{A_{\Delta\alpha^*}^2 + A_{\Delta\delta}^2}$. Although this estimate does not take into account the inclination or spatial orientation, it provides an appropriate order of magnitude constraint for the spatial scale of the system.

\subsubsection{\textit{Gaia} Astrometry}
\label{Gaia Initial}

Since \textit{Gaia} measurements are projected along a scan angle, the sinusoidal periodicity that Lomb-Scargle is based upon is no longer effective at determining an accurate period. To obtain an initial period estimate for \textit{Gaia} data, we implement the general linear periodogram approach developed by \citet{Delisle2022} for scanning space astrometry. 

We start by creating a grid of trial frequencies $1/P$, whose range $[P_{\text{min}},P_{\text{max}}]$ is determined by the users provided prior on $P$. We then follow the procedure outlined in Section 3 of \citet{Delisle2022}, where we compute the $\chi^2$ of our orbital model at each candidate frequency, where each column in matrix $\mathbf{M}_{\text{astro}}$ corresponding to orbital motion is further parameterized with the mean motion $n = 2\pi/P$. We compute the periodogram power $z_{\text{GLS}}(1/P)$  \citep[Eq. 22]{Delisle2022} as the fractional difference in $\chi^2$ between the null model (Section \ref{subsub:null_models}) $\mathcal{H}$ and periodic model $\mathcal{K}(1/P)$.
A specified number of frequency candidates that maximize $z_{\text{GLS}}(1/P)$ provide the initial $P_{\text{init}}$ estimates.

\subsubsection{Radial Velocity}

The periodicity in radial velocity data can also be highly non-sinusoidal, which means that a separate technique from the LS periodogram is required to provide an initial guess on the period. We use the Phase Distance Correlation (PDC) periodogram developed by \citet{ZuckerPDC}, which is a generalization of the Generalized Lomb-Scargle periodogram (GLS) \citep{2009Zechmeister}, capable of finding non-sinusoidal periodicities. The PDC serves as an appropriate framework to seed our initial period estimate, as it performs similarly to GLS in sinusoidal regimes while also handling highly eccentric orbital signals.  

We start by creating a grid of trial periods whose range is set by the users prior on the period $P$. We then calculate the PDC for each trial period using the recipe described in Section 2 of \citet{ZuckerPDC}. The number of specified trial periods that maximize the PDC \citep[Eq. 5]{ZuckerPDC} are then selected as the initial estimates for $P_{\text{init}}$.

\subsubsection{Periodogram for Joint Datasets}

Initialization for joint datasets requires us to find a common periodicity across an arbitrary number of datasets. To accomplish this, we use the framework developed in Section \ref{linearization}, utilizing the design matrix $\mathbf{M}$ to perform a $\chi^2$ grid search for the optimal joint period. Following the approach outlined in \citet[Section 4.2]{Delisle2022}, each column corresponding to orbital motion in $\mathbf{M}$ is further parameterized by the mean motion $n$. We then compute the $\chi^2$ for each trial frequency $1/P$ across a grid bounded by the user's prior on $P$. The $z_{\text{GLS}}$ described in Section \ref{Gaia Initial} is then calculated. The specified number of trial frequency's that maximize the $z_{\text{GLS}}$ yields the inverse of the initial period estimates $P_{init}$.

\subsubsection{Full Seed Creation}

For each initial period estimate obtained from the periodogram methods described above, a candidate parameter vector is created by filling the remaining parameters with a random sample from their priors. These are then each taken to seed \texttt{SciPy}'s Differential Evolution \citep{StornRainer} algorithm, which searches the entire parameter space for a global minimum. The linearized model discussed in Section \ref{linearization} is used in the likelihood for differential evolution to prevent the optimizer from stalling in the large, full parameter space. The linear parameters are then transformed to the desired parameters for sampling.

The returned parameter vector corresponding to the highest likelihood from the differential evolution runs is then sent through a quick minimization routine with \texttt{SciPy}'s minimize. The final parameter vector returned from this process is then chosen as the initial guess. Finally, a small Gaussian ball is created around the initial guess to seed \texttt{emcee}.

\subsection{Sampling}
\label{sampling}

Determining the joint posterior distribution of orbital parameters poses many challenges, including strong degeneracies between parameters and highly non-linear parameter spaces. \texttt{periapsis} utilizes Bayesian posterior sampling to provide robust estimates on orbital parameters and their uncertainties. We offer two different samplers for Bayesian inference: affine invariant Markov Chain Monte Carlo sampling with \texttt{emcee} \citep{emcee} and nested sampling via \texttt{UltraNest} \citep{Ultranest}. These two methods offer different ways of exploring the parameter space and sampling from the posterior distribution. This allows users to analyse their orbital parameters effectively and to ensure that a consistent solution exists across different sampling methods. Here, we describe the sampling methods offered and how they are implemented in \texttt{periapsis}.

\subsubsection{\texttt{emcee}}
\label{subsubsec:emcee}

\texttt{emcee} is an affine invariant Markov Chain Monte Carlo (MCMC) sampler. The sampler evolves an ensemble of starting walkers through the parameter space, where each walker's position is updated based on the probability of the next proposed position. This allows for efficient sampling of the posterior probability density distribution, but is heavily dependent on the initial position of the walkers and can become trapped in the local minima of the distribution. This potential problem is generally resolved by our initial guessing schemes presented in Section \ref{initial guess}. Tracking the convergence of the walkers is important to determine the reliability of the orbital fit and the returned posterior distribution. The integrated autocorrelation time is seen as the best measure of convergence, which accounts for noise in the chain and determines the number of steps required to draw independent samples \citep{emcee}. A shorter autocorrelation time indicates better performance of the sampler. The effective sample size (ESS) is a proxy of the autocorrelation time, as it measures how many independent samples are obtained; the more independent samples, the better the convergence. The acceptance fraction, which is how often a walker step is accepted, also acts as a good indicator for MCMC performance. An acceptance fraction between $0.2-0.5$ typically indicates good performance, although this range is dependent on the dimensionality and geometry of the parameter space \citep{emcee,Li2024}. 

The user provides the amount of walkers and iterations that are desired for the sampler, where typically using a higher amount of walkers is advised \citep{emcee}. Once the sampler is  initialized, there is a period of time before the walker chains become independent of each other, known as the burn-in phase. These samples should be tossed out from the final analysis for robust posterior distribution estimates. We choose to discard two times the maximum autocorrelation time for burn-in, and we thin the samples by two times the minimum autocorrelation time. 

Two separate classes are offered for sampling with \texttt{emcee}, one for sampling the full parameter vector, \texttt{MCMCFitter}, and one for sampling only the non-linear parameters, \texttt{MCMCLinearFitter}. Both classes use the same log likelihood function presented in Equation \ref{eq:lnlike}, but the model positions are computed differently. \texttt{MCMCFitter} uses the models shown in Section \ref{astrometry_model} and \ref{rv_model} to sample each parameter. Fitting on the full parameter vector allows for analysis on the covariances between the parameters and robust estimation on their posterior distributions. However, the large parameter space associated with this leads to longer computational and convergence time.

\texttt{MCMCLinearFitter} is offered as a faster option that only samples the non-linear parameters, using the method described in Section \ref{linearization}. As this method reduces the parameter space, the sampler is able to converge more efficiently. The linear parameters are then optimized for each sampled non-linear one. In order to accurately sample the posterior of each linear parameter and account for covariance between parameters, we perform a QR factorization on the matrix $\mathbf{WM}$ where the matrix is factored into an orthonormal part $\mathbf{Q}$ and an upper triangular part $\mathbf{R}$ such that $\mathbf{QR} = \mathbf{WM}$ \citep{Francis1961}. We then sample by solving (by back-substitution)

\begin{equation}
    R\vec{\mu}_{sampled} = \mathbf{Q}^T\mathbf{W}\vec{x} + \mathbf{\epsilon}
\end{equation}

where $\mathbf{\epsilon} \sim \mathcal{N}(0,1)$ is Gaussian noise. This step is inspired by the procedure described in \citet{Bardsley2014} and accurately recovers the posterior distribution of the linear parameters found by bootstrapping or empirically calculating the covariance matrix of the marginalized linear parameters at each sampling step as in \citet{TheJoker2017}.

A numerical bootstrapping procedure using the linear matrix is also provided in the package in case users would like to perform a bootstrap to the data by sampling with replacement and performing a linear fit on the resampled data to get another estimate of the uncertainty in each parameter.

\subsubsection{\texttt{UltraNest}}
\label{subsubsec:ultranest}

In situations where the posterior distribution is multi-modal, or in cases of sparse data and poor initial estimates, \texttt{UltraNest} \citep{Ultranest} provides a robust alternative to MCMC samplers like \texttt{emcee}. \texttt{UltraNest} is a nested sampling algorithm, which takes a set of $N$ points from a unit hypercube defined by the priors (Section \ref{priors}) and evaluates the likelihood for each of them. The lowest likelihood point is removed and replaced by a point within the newly constrained prior volume. This iteratively shrinks the investigated prior space, leaving each weighted “dead point” as a posterior sample. This process stops when the remaining live points have negligible contribution to the total estimated evidence. \texttt{UltraNest} uses reactive nested sampling \citep{Ultranest}, which allows for exploration of the entire parameter space as it begins sampling from the entire prior volume. This is a reliable method for exploring difficult parameter spaces without getting trapped in local minima, which can be a common issue for \texttt{emcee}.

\texttt{UltraNest} offers many helpful features for tuning convergence and parameter exploration. Orbital parameters that are periodic in nature (such as angles or phases spanning $2\pi$) are specified as wrapped parameters inside the sampler, which prevents artificial discontinuities across periodic boundaries during nested sampling. Users can specify thresholds on the minimum number of live points sampled, the target effective sample size, and the evidence tolerance \texttt{dlogz}. These parameters control the accuracy of the posterior and the computational cost for likelihood evaluations. For example, a smaller \texttt{dlogz} requires the sampler to continue exploring the parameter space until the remaining contribution to the evidence falls below a tighter threshold. By default, \texttt{dlogz} = 0.5, which allows the sampler to converge quickly since we do not place a strict evidence tolerance. For poorly converged solutions, we recommend decreasing this threshold.   
  
As with \texttt{emcee}, two separate classes are offered for sampling with \texttt{UltraNest}, \texttt{UltranestFitter} and \texttt{UltranestLinearFitter}. \texttt{UltranestFitter} samples the full parameter vector, calculating the model positions using the framework in Sections \ref{astrometry_model} and \ref{rv_model}. Exploring the full parameter space with \texttt{UltranestFitter} can be computationally expensive and leads to long run times, so we generally advise to utilize nested sampling with the \texttt{UltranestLinearFitter} class, which reduces the parameter space down to only the non-linear parameters, using the same logic as \texttt{MCMCLinearFitter} (Section \ref{subsubsec:emcee}). This allows the sampler to converge very quickly while scanning the full parameter space for multi-modalities, acting as an important validation for avoiding local minima. The linear parameters are also sampled via $\mathbf{QR}$ decomposition, as described in Section \ref{subsubsec:emcee}.

\subsection{Output}
\label{Output}

\texttt{periapsis} offers many ways to analyse the fitted data after sampling. 
The results object returned after sampling is a comprehensive output featuring the full samples, the maximum a posteriori (MAP) and median parameters, initial period guesses from the periodogram (Section \ref{initial guess}), and other information from the fitting procedure. Parameters that are derivable from the sampled parameters can be retrieved from the results object in any unit supported by the package. The median parameters and their uncertainties are returned in the user supplied units, while the MAP and median parameter vectors are returned in the canonical units (Section \ref{data and units}) so that they can easily be supplied to the \texttt{Orbit} class (Section \ref{subsec:synthetic}).    

Goodness of fit metrics and statistics on the performance and convergence of the sampler can be obtained from the statistics suite. The \texttt{all\_stats} function returns all offered statistics, including reduced $\chi^2$ values for the MAP and median fits, the $\Delta\chi^2$ between the orbital fit and the null hypothesis, and other goodness of fit statistics. Convergence measuring quantities like the ESS and integrated autocorrelation time are also returned from this function.

A large number of plots are offered from the plotting suite, providing publication ready visualizations for posterior parameter distributions, convergence diagnostics, and orbital fits. This removes the need to use multiple packages to perform Bayesian inference and then visualize the fit, particularly for \textit{Gaia} scanning plane astrometry.

\subsection{Synthetic Binary Creation}
\label{subsec:synthetic}

\texttt{periapsis} can also be used to create synthetic binary systems for testing. This capability is provided through the \texttt{Orbit} class, which is also used in the likelihood for the full parameter fitter classes. By supplying a parameter vector to the \texttt{Orbit} class, the user can generate model positions or velocities, to which observational errors can be applied to create a synthetic binary. These synthetics are used throughout the \texttt{periapsis} testing suite to analyse the sampler performance and ensure that the true orbital parameters are recovered.

\section{Experiments}
\label{sec:experiments}

To demonstrate the varied features of {\tt periapsis}, in this section we apply it to varied published binary data sets.  This enables us to both check its output against published values, and examine the impacts of varied sampling and fitting methods.  Each example in this section is accompanied by a python notebook available on GitHub \footnote{\url{https://github.com/matthewwhitaker/periapsis_examples}}.

\subsection{oMEGACat BH-2}
The code was developed in the context of fitting binaries in $\omega$~Cen.  Our first example comes from the first binary published from that data set in \citet{Whitaker2026}, oMEGACat BH-2.  The fits presented in that paper were from an earlier version of the code, and thus we don't expect identical results. The dataset includes 363 2D proper motion data points from the Hubble and James Webb Space Telescope spanning 23 years after rejection of outliers in the HST data.  We fit the data in two ways; using linearized parameters and sampling with {\tt UltraNest}, and fitting all parameters using {\tt emcee} as a sampler.  

We use simple uniform priors with a maximum period of 200 years in both fits and achieve quite comparable fits to those presented in \citet{Whitaker2026}, which used similar priors, but implemented a velocity dispersion based prior on the proper motions. Both the \texttt{UltraNest} and \texttt{emcee} fits find a reduced $\chi^2$ of 1.65, identical to that quoted in the original paper, with the difference between reduced $\chi^2$ on the order of 0.01. The best-fit period and eccentricity is directly comparable in both fits.  Our {\tt Ultranest} fit give a period of 90$^{+60}_{-40}$ years, while the {\tt emcee} fit gives 100$^{+60}_{-50}$ years, quite comparable to the 94$^{+63}_{-42}$ years in the paper.  The eccentricity is 0.70$^{+0.09}_{-0.13}$ in the {\tt Ultranest} fit, 0.72$^{+0.07}_{-0.14}$ in {\em emcee} and 0.72$^{+0.08}_{-0.13}$ in the paper.  Finally, the inferred companion mass  (assuming an 0.78~M$_\odot$ primary mass in all cases) is 4.44$^{+1.22}_{-1.05}$~M$_\odot$ in the {\tt Ultranest} fit, 4.60$^{+1.30}_{-1.02}$~M$_\odot$ in the {\tt emcee} fit, and 4.46$^{+1.22}_{-1.01}$~M$_\odot$ in the paper, showing quite consistent agreement despite the slight varying priors.  In general, the {\tt Ultranest} fits provide much smoother looking parameter distributions due to the lower dimensionality of the fits and the lack of sampling in the linearized parameters, but the period-eccentricity distributions look almost identical, showing strong covariance.

\subsection{Gaia BH3}
\begin{figure*}
\includegraphics[width=\textwidth]{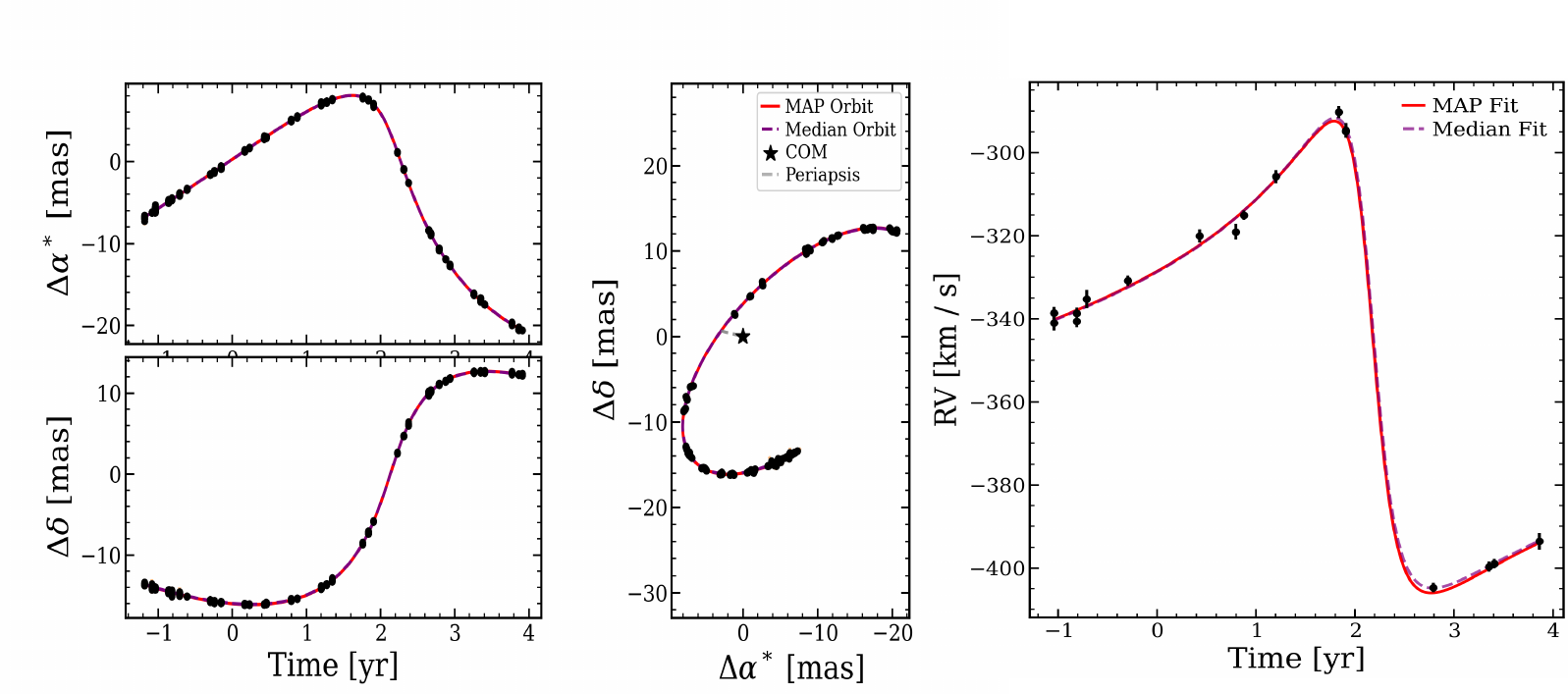}
\caption{Orbital solution from \texttt{periapsis} for Gaia BH3 fit to data from \citet{GaiaBH3_2024}, demonstrating some of the plots produced by the package.  Left panels show astrometry timelines after removing the best-fit proper motion; these are combined to show the orbit on the sky in the middle panel.  The right panel shows the radial velocity data.  In all panels, the maximum a posteriori (MAP) fits and median fits are shown as solid red and dashed purple lines; in this case they are indistinguishable.}
\label{fig:gaiabh3}
\end{figure*}

Gaia BH3 is also long-period visible star--black hole binary, with a published period of 11.6 years, and an inferred dark companion mass of 32.70$\pm$0.83~M$_\odot$ \citep{GaiaBH3_2024}.  The dataset available is different from oMEGACat BH-2, comprising 599 unflagged Gaia astrometric measurements and 17 Gaia radial velocity measurements both spanning 5 years.  The \citet{GaiaBH3_2024} paper presents two solutions, both with astrometry and a joint fit with astrometry and kinematics.  The \texttt{periapsis} fit of this data is consistent with the presented values with a best-fitting mass for both solutions of $33.3^{+0.6}_{-0.6}~\mathrm{M_{\odot}}$.  We compare the combined fits here, but the astrometric only fit is also available in the example notebook, as is a separate notebook example where the DR4 data is downloaded and fit; each of these give somewhat different answers.  The joint fit using \texttt{emcee} yields a reduced $\chi^2$ of 0.92, with a best-fitting period of ${P} = 11.59^{+0.26}_{-0.25}~\mathrm{yr}$, an eccentricity of ${e} = 0.728^{+0.005}_{-0.005}~\mathrm{}$, and a semi-major axis of ${{{a}_{1}}_{\mathrm{ang}}} = 27.3^{+0.5}_{-0.5}~\mathrm{mas}$; these all are consistent with errors of the best-fit joint solution published in \citet{GaiaBH3_2024}.  

We also use Gaia BH3 to provide examples of plots that are produced using the \texttt{periapsis' all\_plots} command in Fig.~\ref{fig:gaiabh3}. We show the orbital fit plots here, but this command also produces parameter autocorrelation and effective sample size plots to examine convergence, corner plots and 1D histograms of all fitted parameters, residual plots, and inferred companion mass histograms.

\subsection{NGC 3201}

Three candidate star--black hole binaries have been reported via untargeted MUSE spectroscopy of the low-density Galactic globular cluster NGC 3201 by \citet{Giesers2018,Giesers2019}. Here we re-fit the existing radial velocities for these three candidates.

N3201-BH1 (ID \#2560) was the first of these black holes reported \citep{Giesers2018}, with 35 published radial velocities spanning more than four years. We first fit with uniform, uninformative priors using \texttt{emcee}, along with the published assumed visible star mass of $0.81\pm0.05 M_{\odot}$. We find essentially identical values of all orbital parameters to those published in \citet{Giesers2019} and the same minimum black hole mass of $M \sin{i} = 4.53\pm0.15 M_{\odot}$, though our uncertainty is a bit smaller than the published uncertainty of $\pm0.21 M_{\odot}$. Adding a prior based on the systemic velocity and velocity dispersion of the globular cluster (mean 494.0 km s$^{-1}$, $\sigma = 5.4$ km s$^{-1}$) does not change the inferred parameter values within their reported uncertainties, because the data already constrain the velocity of the binary to a value near the cluster mean to high precision ($\gamma = 494.3\pm0.4$ km s$^{-1}$).

N3201-BH2 (ID \#21859) is the second convincing black hole previously identified. We refit the eight radial velocities published in \citet{Giesers2019} with uniform priors using \texttt{emcee}, along with the published assumed visible star mass of $0.61 M_{\odot}$. We find a bimodal posterior distribution, with the more prominent peak consistent with the parameters found in \citet{Giesers2019}, and a secondary region of the posterior centred on a systemic velocity of about 360 km s$^{-1}$, inconsistent with the cluster. If we change the prior on $\gamma$ to the Gaussian prior described above ($494.0 \pm 5.4$ km s$^{-1}$) and make no other changes, many of the \texttt{emcee} walkers still never reach the tiny phase space of the global minimum, but setting the orbital period prior to values within 10\% of MAP value is sufficient to address the issue. This challenge highlights the sensitivity of the fitting process to the priors when $N$ is low (see additional discussion in Section \ref{sec:usage}). Additionally assuming the published visible star mass prior of $0.61\pm0.05 M_{\odot}$, our final minimum black hole mass
is $M \sin{i} = 7.53^{+0.27}_{-0.25} M_{\odot}$, consistent with but more precise than the previous value of $7.68\pm0.50 M_{\odot}$. 

While \citet{Giesers2019} only publish eight velocities for this binary, three additional velocities that fell below their nominal spectroscopic signal-to-noise cut are plotted in their Figure A.1. Adding these three unpublished values and refitting changes the minimum black hole mass only very slightly to 
$M \sin{i} = 7.58^{+0.27}_{-0.26}~\mathrm{M_{\odot}}$, and all other parameters likewise match the $N=8$ fits well within their uncertainties. 

We note in passing that N3201-BH2 shows H$\alpha$ emission, which was suggested as evidence for ongoing accretion by \citet{Giesers2019}. Since the donor is far from filling its Roche lobe for the best-fit binary parameters, it is much more likely that the H$\alpha$ emission reflects enhanced chromospheric activity due to the visible star being spun up from tidal synchronization to the black hole at its short 2.2 d period.


A third binary in NGC 3201, ID \#5132, is a possible black hole candidate with 22 published velocities. \citet{Giesers2019} report a bimodal posterior, with the more probable peak at ($P \sim 760$ d, $e \sim 0.3$) and the less probable peak at ($P \sim 470$ d, $e \sim 0.94$). The more probable peak corresponds to a minimum mass $M \sin{i} = 4.4\pm2.8 M_{\odot}$.

When we refit these data using a uniform prior fit with \texttt{emcee}, together with the published visible star mass of $0.64 \pm 0.05 M_{\odot}$, we find that the posterior is exceptionally sensitive to the exact prior limits used, and that we can get a posterior dominated by each of these solutions with only slight variations in the assumptions. In particular, our default eccentricity prior for this fit is bounded above at $e=0.95$ for numerical stability, as the fits become unstable as $e$ approaches 1. Since the high-eccentricity solution sits near this boundary, the allowed parameter space volume is truncated and therefore not strictly uninformative. For this reason we cannot assign a relative probability to these two solutions with the existing data.

By appropriate adjustments of the prior, we can get stable posteriors that capture each of these two families of possible solutions. Additionally assuming the globular cluster $\gamma$ prior, for the higher-mass solution, we find a minimum mass $M \sin{i} = 4.2^{+2.0}_{-1.4}~\mathrm{M_{\odot}}$, consistent with that of \citet{Giesers2019} but with somewhat smaller uncertainties. This solution has a reduced $\chi^2 = 1.29$. The lower-mass solution has a minimum mass $M \sin{i} = 1.23^{+0.52}_{-0.27}~\mathrm{M_{\odot}}$, which would still be consistent with a compact object. Formally this solution has a lower reduced $\chi^2$ of 1.04, placing the single most outlying velocity measurement near periastron. As few as one or two additional velocity measurements made near this phase could confirm or rule out this solution and hence potentially confirm another black hole in NGC 3201.

We also explored whether an \texttt{UltraNest} fit would find a bimodal posterior. It did, but barely. There were very few posterior samples in the lower-mass, high eccentricity solution compared to what was found in \citet{Giesers2019}, emphasizing the sensitivity of the conclusions to the prior.

\subsection{51 Peg b}

51 Peg b was the first exoplanet ever discovered orbiting a Sun-like star \citep{Mayor1995}, with an orbital period of $4.23$ days and an inferred minimum planetary mass of $0.472 \pm 0.039\,M_{\mathrm{Jup}}$ \citep{Butler2006}. We fit this system using 46 high-precision radial velocity measurements from the HIRES spectrograph on the 10-m Keck-1 telescope spanning roughly $8.2$ years published by \citet{Tal-Or2019}, who provided a corrected version of the catalog produced by \citet{Butler2017}. \citet{Butler2017} only provides published orbital parameters for  $P, K, \sigma_{\mathrm{jit}}$ and $\dot{\gamma}$ for 51 Peg b, so we compare both to this text and the full Keplerian parameters published in \citet{Butler2006}. 

We first perform an orbital fit by sampling only the non-linear parameters period $P$, eccentricity $e$, and the scaled time of periastron $t_0$ with \texttt{emcee}. Noninformative uniform priors are placed on $P$ and $t_0$, while a log uniform prior is placed on $e$, and the star's mass is fixed to the published value $1.09\, M_{\odot}$. We obtain nearly identical estimates of $P$ and $K$ as \citet{Butler2017}, and a similar planetary mass of $M \sin{i} = 0.478 \pm 0.002\,M_{\mathrm{Jup}}$, having smaller uncertainties but comparable to \citet{Butler2006}. We obtain an eccentricity of $e= 0.021 \pm 0.004$, slightly higher than that published in \citet{Butler2006} of $e = 0.013 \pm 0.012$. In this fit, we did not take into account any stellar activity with a radial velocity jitter term $\sigma_{\mathrm{jit}}$ (Section \ref{subsub:jitter}), so the fit's reduced $\chi^2$ is significantly inflated.

We then fit a full parameter vector, including $\sigma_{\mathrm{jit}}$ and $\dot{\gamma}$, with \texttt{emcee}. We again obtain nearly identical values to the $P$ and $K$ published by \citet{Butler2017}, and an $M \sin{i} = 0.476 \pm 0.005\, M_{\mathrm{Jup}}$, again with smaller uncertainties but comparable to \citet{Butler2006}. We obtain a similar eccentricity as the last fit, but with larger uncertainties where $e = 0.020 \pm 0.011$, comparable to those published by \citet{Butler2006}. The estimates for $\dot{\gamma} = -0.11 \pm 0.20~\mathrm{m\,s^{-1}\,yr^{-1}}$ and $\sigma_{\mathrm{jit}} = 2.67^{+0.40}_{-0.33}~\mathrm{m\,s^{-1}}$ differ from $\dot{\gamma} = 0.287 \pm 0.247~\mathrm{m\,s^{-1}\,yr^{-1}}$ and $\sigma_{\mathrm{jit}} = 2.16 \pm 0.36~\mathrm{m\,s^{-1}}$ published by \citet{Butler2017} because we fit the corrected catalog provided by \citet{Tal-Or2019}. Sampling for jitter remedies the inflated reduced $\chi^2$, yielding a reduced $\chi^2= 1.02$. 




\section{Discussion}
\label{sec:discussion}
\subsection{Comparison to Existing Packages}

There are several other orbital fitting codes. Many are useful in specific cases, especially for multi-body exoplanet systems, but to our knowledge, none have both the flexibility and extensibility of \texttt{periapsis} for fitting two-body systems. Here we briefly describe these other codes.

\subsubsection{\textit{The Joker}}

\textit{The Joker} (a portmanteau of Johannes Kepler) is a Python rejection sampling code that supports radial velocity data \citep{TheJoker2017}. \textit{The Joker} excels with sparse radial velocity data where the posterior is multimodal. Rejection sampling is powerful here due to the number of samples that are tested, often exceeding $10^8$ \citep[e.g. $2^{29} \approx 5\times 10^8$ for each of the binaries tested in NGC 3201;][]{Giesers2019}. The generation of a large number of prior samples is slow (generating $2^{29}$ samples takes about a day on an Apple Silicon M1 Max with 64 GB of RAM), so \textit{The Joker} may be best suited to cases where the same prior samples can be reused for a large number of binaries. In cases when an insufficient number of samples are obtained by rejection sampling, \textit{The Joker} recommends using MCMC sampling to augment the existing samples and explore the nearby posterior distribution. 

\textit{The Joker} directly samples $P$, $e$, $\omega$, and $M_0$ (which they call $\phi_0$). The sampler marginalizes over the linear parameters $K$ and $\gamma$ (which they call $v_0$). \textit{The Joker} supports \texttt{astropy}'s Quantity for input data and priors so arbitrary units are allowed. 

\subsubsection{\texttt{kepmodel}}

\texttt{kepmodel} is an analytical periodogram-based orbital fitting code optimized for the \textit{Gaia} pipeline \citep{Delisle2022}. A candidate period is first identified from joint 1D scan-plane astrometry and radial velocities by a linear periodogram, which we replicate in our initial guess scheme as described above (Section \ref{Gaia Initial}). \texttt{kepmodel} then analytically estimates $e$ and $M_0$ based on the amplitudes and phases from the periodogram. With these estimates, one can solve for the linear parameters $A, B, F, G$, $K\sin\omega,$ and $K\cos\omega$.

In practice, \texttt{kepmodel} is more flexible than described here, allowing users to add an arbitrary number of Keplerian models to a single system (e.g. for multi-planet fits) and a flexible \texttt{S+LEAF} noise model \citep{Delisle2020}. The user can define arbitrary linear parameters in addition to the orbital parameters described above, including parallax, acceleration terms, and instrument offsets. Although \texttt{kepmodel} fits data using the Thiele Innes parameterization, transformations to the following Keplerian parameters (as defined in Table \ref{tab:periapsis_params}) and their combinations are supported: $n$, $P$, $\log_{10}P$, $M_0$, $M_0 + \omega$, $L_0$, $T_p$, $T_c$, $T_{v_\text{min}}$, $T_{v_\text{max}}$, $K$, $\log_{10}K$, $a_1$, $a_1\sin i$, $e$, $\omega$, $e\cos\omega$, $e\sin\omega$, $\sqrt{e}\cos\omega$, $\sqrt{e}\sin\omega$, $\tilde{\omega}$, $i$, and $\Omega_{asc}$.

\subsubsection{\texttt{kima}}

\texttt{kima} is a C++ code with a Python interface that uses diffusive nested sampling \citep{Faria2018,Baycroft2026}. Primarily designed for exoplanet fitting, the defining feature of \texttt{kima} is its ability to fit planetary systems where the number of planets is not known \textit{a priori}. A number of other models are also supported, including gaussian process terms and binary interaction terms. Recently \texttt{kima} was expanded to support \textit{Gaia} astrometry \citep{Baycroft2026}.

\subsubsection{\texttt{Octofitter}}

\texttt{Octofitter} is a Julia code primarily designed for fitting orbits of directly imaged exoplanets \citep{Thompson2023}. It is named for the eight types of data it supports: relative astrometry, absolute and/or relative star and planet radial velocities, \textit{Hipparcos}-\textit{Gaia} proper motion anomaly, images, and interferometric data. For sampling, \texttt{Octofitter} uses a No U-Turn Hamiltonian Monte Carlo Sampler. The package can also sample from a wide range of orbital parameter bases.

\subsubsection{\texttt{orbitize!}}

\texttt{orbitize!} is a Python package based on the Orbits For the Impatient \citep[OFTI;][]{Blunt2017} algorithm, with support for MCMC and nested sampling as well \citep{Blunt2020,Blunt2024}. OFTI is an optimized rejection sampling algorithm that is ideal when the data covers only a small fraction of the orbital period. \texttt{orbitize!} allows users to perform sampling in a number of predefined parameter bases, with the standard basis defined as $a$, $e$, $i$, $\omega$, $\Omega$, $t_0$ (their $\tau$), $\varpi$, and $M_{tot}$.

\subsubsection{\texttt{orvara}}

Orbits from Radial Velocity, Absolute, and/or Relative Astrometry (\texttt{orvara}) is a Python package that supports the data types listed in its name \citep{Brandt2021}. The posterior distribution is sampled using MCMC. Sampling is done on $M_1$, $M_2$, $a$, $\sqrt{e}\cos\omega$, $\sqrt{e}\sin\omega$, $i$, $\Omega_{asc}$, $L_0$, $\varpi$.

\subsection{Usage Recommendations: How to get a quick orbital solution}
\label{sec:usage}

\texttt{periapsis} offers many ways to characterize a binary system with Bayesian inference. The variety of methods offered allows users to confidently validate their orbital fits on a variety of datasets, big and small. Here, we offer our recommendations on how to quickly obtain and validate orbital solutions for any system of interest.

In most cases, we believe it is best to first perform an orbital fit with the \texttt{UltranestLinearFitter} class (Section \ref{subsubsec:ultranest}), which can efficiently scan the parameter space of the non-linear parameters. This is important to identify potential multimodality within the parameter space that can impact the orbital fit. For validation of the fit and a more in-depth analysis of the orbital parameters, we then recommend fitting with the \texttt{MCMCFitter} class (Section \ref{subsubsec:emcee}). This samples the entire orbital parameter vector, exploring the full parameter space rather than optimizing the linear ones. Using both fitting methods on the data provides a way to validate the best-fitting solution and offers different ways of analyzing the sampled parameters. The full fit method is computationally more expensive, so for large binary datasets, it is best to do a first run through with one of the linear methods, and then analyse interesting systems thoroughly with \texttt{MCMCFitter}. 

\texttt{UltranestLinearFitter} can struggle to converge for sparse datasets ($N \lesssim 10$), potentially leading to long computational times. In this case, we recommend using the \texttt{MCMCFitter} or \texttt{MCMCLinearFitter} classes. Both methods utilize \texttt{emcee}, which is seeded with our periodogram based initialization methods presented in Section~\ref{initial guess}, where a differential evolution run searches around candidate periodogram peaks before seeding \texttt{emcee}. The user can select the number of peaks used for initialization with the \texttt{npeaks} argument. In the sparse data regime, we recommend setting \texttt{npeaks} $\geq 5$, which can help the sampler locate high probability solutions that sit in narrow modes of the parameter space.

\subsection{Future Use Cases}

The upcoming several years shows great promise for significant progress in our knowledge of binary stars.  Astrometrically, the coming release of Gaia Data Release 4 and the upcoming expected high-precision astrometric data from the Roman Space Telescope \citep[e.g.][]{Lam2026} are both expected to find many binaries.  The flexibility of \texttt{periapsis} to handle both these astrometric data and any additional follow up data will enable efficient orbital fitting of new sources.  Ongoing and upcoming spectroscopic surveys such as 4MOST \citep{deJong2019} will also enable discovery of binaries with several dedicated subsurveys devoted to finding binaries.

\section*{Acknowledgements}

EK, MW, and AS acknowledge support from  Hubble Space Telescope
archival grant HST-AR-17033 and James Webb Space Telescope general observer grant JWST-GO-5137. JS acknowledges support from NASA grant 80NSSC26K0226.
We would like to thank Kameron Goold for assistance in naming \texttt{periapsis}, and Gail Zasowski for helping test the package. The support and resources from the Center for High Performance
Computing at the University of Utah are gratefully
acknowledged.

During software development for \texttt{periapsis}, generative AI was used occasionally to provide debugging and coding assistance and as a backboard for proofreading code and text written by the authors. All prose, mathematical formulations, code, and scientific conclusions were written, derived, and independently verified by the authors.

\section*{Data Availability}

The \texttt{periapsis} code presented in this paper is available at \url{https://github.com/Evank2004/periapsis}, and example notebooks containing the analysis in Section~\ref{sec:experiments} are available at \url{https://github.com/matthewwhitaker/periapsis_examples}. The observational datasets analysed in this work all come from previously published studies: 2D astrometry from the \emph{Hubble Space Telescope} and the \emph{James Webb Space Telescope} for oMEGACat BH-2 from \citet{Whitaker2026}; astrometry and radial velocities for Gaia BH3 from \citet{GaiaBH3_2024}; MUSE radial velocities for the NGC 3201 candidates from \citet{Giesers2018,Giesers2019}; and the 51 Peg b radial velocities from \citet{Tal-Or2019} (improving on \citet{Butler2017}).

\bibliographystyle{mnras}
\bibliography{main}

\appendix 

\section{Supported Parameters and Transformations} 
\label{app:sec1}

\texttt{periapsis} supports a large variety of parameters for sampling, presented in Table \ref{tab:periapsis_params}. Below, we include the complete relationships between parameters not included in Section \ref{sec:Keplerian Models}.

\subsection{Thiele-Innes and Velocity Constants}
\label{app:Thiele-Innes}
The Thiele-Innes constants are related to the Campbell elements $a$, $i$, $\omega$ and $\Omega$ by: 
\begin{align}
    \label{eq:thieleinnes_from_campbellA}
    A &= a (\cos\Omega \cos\omega - \sin\Omega \sin\omega \cos i), \quad A_1, A_2 \\
    \label{eq:thieleinnes_from_campbellB}
    B &= a (\sin\Omega \cos\omega + \cos\Omega \sin\omega \cos i), \quad B_1, B_2 \\
    \label{eq:thieleinnes_from_campbellF}
    F &= a (-\cos\Omega \sin\omega - \sin\Omega \cos\omega \cos i), \quad F_1, F_2 \\
    \label{eq:thieleinnes_from_campbellG}
    G &= a (-\sin\Omega \sin\omega + \cos\Omega \cos\omega \cos i), \quad G_1, G_2 \\
    \label{eq:thieleinnes_from_campbellC}
    C &= a \sin i \sin\omega, \quad C_1, C_2 \\
    \label{eq:thieleinnes_from_campbellH}
    H &= a \sin i \cos\omega, \quad H_1, H_2
\end{align}
To transform from the Thiele-Innes constants to the Campbell elements, we follow the procedure outlined in \citet{Popovic1995}.

Conversions between velocity constants ($c, h$) and semi-amplitudes ($K$) are:
\begin{align}
    c &= -K \sin\omega, \quad c_1, c_2 \\
    h &= K \cos\omega, \quad h_1, h_2 \\
    K &= \sqrt{c^2 + h^2}, \quad \omega = \arctan2(-c, h)
\end{align}

\subsection{Distances and Ellipse Geometry}
\label{app:dist_geo}
Transformations between semi-major axis $a$, semi-minor axis $b$, semilatus rectum $p$, periapsis distance $r_{\mathrm{p}}$, and apoapsis distance $r_{\mathrm{a}}$:
\begin{align}
    b &= a \sqrt{|1 - e^2|} \\
    p &= a (1 - e^2) \\
    r_{\mathrm{p}} &= a(1 - e), \quad r_{\mathrm{a}} = a(1 + e) \\
    e &= \frac{r_{\mathrm{a}} - r_{\mathrm{p}}}{r_{\mathrm{a}} + r_{\mathrm{p}}}, \quad a = \frac{r_{\mathrm{a}} + r_{\mathrm{p}}}{2}
\end{align}
Angular astrometric measurements ($a_{\text{ang}}$) convert to physical linear scales using distance $d$ or parallax $\varpi$ (Section \ref{data and units}):
\begin{align}
    a_{\text{phys}} &= a_{\text{ang}} \cdot d = \frac{a_{\text{ang}}}{\varpi}
\end{align}
where distance and parallax are inversely related by $d = 1 / \varpi$ (with $d$ in parsecs and $\varpi$ in arcseconds).

\subsection{Anomalies, Mean Longitudes, and Epoch Conversions}
\label{app:anomalies}
Relationships for the longitude of periastron $\tilde{\omega}$, mean motion $n$, scaled time of periastron $t_0$, mean anomaly relative to reference epoch  $M_0$, mean longitude $L_0$, eccentric anomaly $E_0$, and the true longitude $l_0$:
\begin{align}
    \tilde{\omega} &= \omega + \Omega_{\text{asc}}, \quad \tilde{\omega}_1, \tilde{\omega}_2 \\
    n &= \frac{2\pi}{P} \\
    t_0 &= \frac{T_{\mathrm{p}} - T_{\mathrm{epoch}}}{P} \\
    \label{eq:L_0}
    L_0 &= \tilde{\omega} + M_0 \\
    M_0 &= E_0 - e \sin E_0 \\
    E_0 &= \arctan\left(\frac{\sin\nu_0 \sqrt{1-e^2}}{e + \cos\nu_0}\right) \\
    \l_0 &= \nu_0 + \tilde{\omega}
\end{align}

\subsection{Transformation Graph and Algorithm}

Keplerian orbits can be represented by a large number of parameterizations (see Table \ref{tab:periapsis_params}). These parameterize orbital ellipses, reference times and directions, instrument systematics, and physical observables.

While other packages restrict users to a specific parameterization, \texttt{periapsis} is extremely flexible: for example, a user can define priors on any combination of parameters and then sample the posterior using whatever parameterization they desire, so long as the parameters are sufficient to describe the data. The resulting posterior distribution can be transformed to any parameter defined by the combination of sampled parameters.

To accomplish this, we define a graph where each parameter in Table \ref{tab:periapsis_params} is a node. Edges are directional and map one or more nodes to another one or more nodes. For example, one edge maps $a$, $\Omega$, $\omega$, $\cos i$ to $A$, $B$, $F$, $G$ via equations \ref{eq:thieleinnes_from_campbellA}--\ref{eq:thieleinnes_from_campbellG}. This many-to-many hypergraph can then be traversed to find the transformations necessary to transform a set of known parameters to a set of desired parameters.

The transformations are constructed as follows (for specific implementation details, see the publicly available code):
\begin{enumerate}
    \item Determine the initial set of known parameters $X_i$ and the set of desired parameters $X_f$.
    \item If $X_f \subseteq X_i$, no transformation is needed.
    \item For every parameter reachable from $X_i$, find a short sequence of transformations through the graph (iteratively preferring sequences with fewer steps) from any subset of $X_i$ to that parameter.
    \item Remove all except the sequences which result in the parameters in $X_f$.
    \item Merge the selected sequences into a single plan, removing duplicate transformations and any transformation with results that are all already computed by an earlier step.
\end{enumerate}

This greedy algorithm quickly finds a small number of transformations to reach $X_f$ from $X_i$, although it does not guarantee the smallest possible number of transformations, which is computationally hard to find. Additionally, once the transformation plan $X_i \rightarrow X_f$ is constructed, it is cached in memory and reused, so that subsequent calls to compute the plan come at negligible cost.

By construction from the graph, the plan is already in a topological order, such that any transformation that depends on others is completed after all those others have finished. So, running the plan only requires calling each transformation in the plan in order. These cached transformation plans allow us to efficiently transform from one basis to another in every likelihood call in order to evaluate priors and constraints on parameters that are not being directly sampled. For example, a user could put a constraint on $a$, while directly sampling only $M_\mathrm{tot}$ and $P$ (Section \ref{priors}).

\begin{table*}
\centering

\caption{Complete table of parameters for \texttt{periapsis}. The Core column indicates parameters solved analytically via linear least squares or sampled directly in the standard model by default. Italicized text in the Code Notation column represents input by the user.}
\label{tab:periapsis_params}
\begin{tabular}{p{5.5cm} p{3.5cm} p{4.0cm} c}
\hline
Description / Name & Parameter & Code Notation & Core \\
\hline
Orbital period & $P$ & \texttt{P} & $\checkmark$ \\
Eccentricity & $e$ & \texttt{e} & $\checkmark$ \\
Time of periastron passage & $T_{\mathrm{p}}$ & \texttt{Tp} & $\checkmark$ \\
Semi-major axis (rel., 1, 2) & $a, a_1, a_2$ & \texttt{a}, \texttt{a1}, \texttt{a2} & \\
Inclination & $i$ & \texttt{i} & \\
Cosine of inclination & $\cos i$ & \texttt{cosi} & \\
Sine of inclination & $\sin i$ & \texttt{sini} & \\
Argument of periastron (rel., 1, 2) & $\omega, \omega_1, \omega_2$ & \texttt{omega}, \texttt{omega1}, \texttt{omega2} & \\
Longitude of the nodes  & $\Omega$ & \texttt{Omega}, \texttt{Omega1}, \texttt{Omega2} & \\
Longitude of the ascending node  & $\Omega_{\mathrm{asc}}$ & \texttt{Omega\_asc} & \\
Parallax & $\varpi$ & \texttt{parallax} & $\checkmark$ \\
Astrometric reference offsets & $\Delta\alpha_0^*, \Delta\delta_0$ & \texttt{dalpha}, \texttt{ddelta} & $\checkmark$ \\
Proper motions & $\mu_\alpha, \mu_\delta$ & \texttt{mu\_alpha}, \texttt{mu\_delta} & $\checkmark$ \\
Proper motion accelerations & $\dot{\mu_{\alpha^*}}, \dot{\mu_\delta}$ & \texttt{dmu\_alpha}, \texttt{dmu\_delta} & \\
Thiele-Innes constants A (rel., 1, 2) & $A, A_1, A_2$ & \texttt{A}, \texttt{A1}, \texttt{A2} & $\checkmark$ \\
Thiele-Innes constants B (rel., 1, 2) & $B, B_1, B_2$ & \texttt{B}, \texttt{B1}, \texttt{B2} & $\checkmark$ \\
Thiele-Innes constants C (rel., 1, 2) & $C, C_1, C_2$ & \texttt{C}, \texttt{C1}, \texttt{C2} & \\
Thiele-Innes constants F (rel., 1, 2) & $F, F_1, F_2$ & \texttt{F}, \texttt{F1}, \texttt{F2} & $\checkmark$ \\
Thiele-Innes constants G (rel., 1, 2) & $G, G_1, G_2$ & \texttt{G}, \texttt{G1}, \texttt{G2} & $\checkmark$ \\
Thiele-Innes constants H (rel., 1, 2) & $H, H_1, H_2$ & \texttt{H}, \texttt{H1}, \texttt{H2} & \\
Thiele-Innes velocity constants c (rel., 1, 2) & $c, c_1, c_2$ & \texttt{c}, \texttt{c1}, \texttt{c2} & $\checkmark$ \\
Thiele-Innes velocity constants h (rel., 1, 2) & $h, h_1, h_2$ & \texttt{h}, \texttt{h1}, \texttt{h2} & $\checkmark$ \\
RV semi-amplitude (rel., 1, 2) & $K, K_1, K_2$ & \texttt{K}, \texttt{K1}, \texttt{K2} & \\
Systemic radial velocity & $\gamma$ & \texttt{gamma} & $\checkmark$ \\
Linear RV trend & $\dot{\gamma}$ & \texttt{dgamma} & \\
Total system mass & $M_{\mathrm{tot}}$ & \texttt{Mtot} & \\
Component masses (body 1, 2) & $M_1, M_2$ & \texttt{M1}, \texttt{M2} & \\
Mass ratio ($M_2/M_1$) & $q$ & \texttt{q} & \\
Mass function (body 1, 2) & $f_1, f_2$ & \texttt{f1}, \texttt{f2} & \\
Minimum component masses & $\min M_1, \min M_2$ & \texttt{minM1}, \texttt{minM2} & \\
Flux ratio & $f$ & \texttt{f\_\textit{band}} & \\
Jitter term & $\sigma_{\mathrm{jit}}$ & \texttt{astro\_jitter}, \texttt{astro\_}\textit{instrument}\texttt{\_jitter}, \texttt{rv\_jitter}, \texttt{rv\_}\textit{instrument}\texttt{\_jitter} &
\\[1.2cm]
Instrument offset & $\mathrm{---}$ & \texttt{astro\_}\textit{instrument}\texttt{\_offset}, \texttt{rv\_}\textit{instrument}\texttt{\_offset}, \texttt{astro\_}\textit{instrument}\texttt{\_x\_offset}, \texttt{astro\_}\textit{instrument}\texttt{\_y\_offset} & \\ 
Reference epoch & $T_{\mathrm{epoch}}$ & \texttt{Tepoch} & $\checkmark$ \\
Semi-minor axis (rel., 1, 2) & $b, b_1, b_2$ & \texttt{b}, \texttt{b1}, \texttt{b2} & \\
Semi-latus rectum (rel., 1, 2) & $p, p_1, p_2$ & \texttt{p}, \texttt{p1}, \texttt{p2} & \\
Apoapsis distance (rel., 1, 2) & $r_{\mathrm{a}}, r_{\mathrm{a}1}, r_{\mathrm{a}2}$ & \texttt{r\_a}, \texttt{r\_a1}, \texttt{r\_a2} & \\
Periapsis distance (rel., 1, 2) & $r_{\mathrm{p}}, r_{\mathrm{p}1}, r_{\mathrm{p}2}$ & \texttt{r\_p}, \texttt{r\_p1}, \texttt{r\_p2} & \\
Longitude of periastron (rel., 1, 2) & $\tilde{\omega}, \tilde{\omega}_1, \tilde{\omega}_2$ & \texttt{piomega}, \texttt{piomega1}, \texttt{piomega2} & \\
Mean motion & $n$ & \texttt{n} & \\
Scaled time of periastron passage ($T_p/P$) & $t_0$ & \texttt{t0} & \\
Mean anomaly at reference epoch & $M_0$ & \texttt{M0} & \\
Mean longitude at reference epoch & $L_0$ & \texttt{L0} & \\
Eccentric anomaly at reference epoch & $E_0$ & \texttt{E0} & \\
True anomaly at reference epoch & $\nu_0$ & \texttt{nu0} & \\
True longitude (epoch, body 1, 2) & $l_0, l_1, l_2$ & \texttt{l0}, \texttt{l01}, \texttt{l02} & \\
Gravitational parameter & $\mu$ & \texttt{mu} & \\
Distance & $d$ & \texttt{distance} & \\
\hline
\end{tabular}
\label{tab:periapsis_params}
\end{table*}

\bsp	
\label{lastpage}
\end{document}